\documentclass[runningheads,a4paper]{llncs}

\usepackage{amssymb}
\usepackage{wrapfig}
\usepackage{amsmath}
\usepackage{graphicx}
\usepackage[noadjust]{cite}
\usepackage{cite}
\usepackage{url}
\usepackage{epstopdf}
\usepackage{bm}
\usepackage{multirow}
\usepackage[table]{xcolor}
\usepackage[font=small,labelfont=bf,tableposition=top]{caption}
\usepackage{subcaption}
\usepackage{xcolor}
\usepackage[bookmarks=false]{hyperref}
\mathchardef\mhyphen="2D
\urldef{\mailsa}\path||
\urldef{\mailsb}\path||
\urldef{\mailsc}\path||    
\newcommand{\keywords}[1]{\par\addvspace\baselineskip
\noindent\keywordname\enspace\ignorespaces#1}

\begin{document}

\title{Location-based Behavioral Authentication Using GPS Distance Coherence}

\author{Tran Phuong Thao}

\authorrunning{T. P. Thao et al.}
%

\institute{University of Tokyo, Japan \\
Graduate School of Information Science and Technology\\
\email{tpthao@yamagula.ic.i.u-tokyo.ac.jp}\\
\email{yamaguchi.rie@i.u-tokyo.ac.jp}\\ $ $ \\
}

\titlerunning{GPS-based Behavioral Authentication Using Distance Coherence}
\maketitle

\begin{abstract}
Most of the current user authentication systems are based on PIN code, password, or biometrics traits which can have some limitations in usage and security.  Lifestyle authentication has become a new research approach. A promising idea for it is to use the location history since it is relatively unique. Even when people are living in the same area or have occasional travel, it does not vary from day to day. For Global Positioning System (GPS) data, the previous work used the longitude, the latitude, and the timestamp as the features for the classification. In this paper, we investigate a new approach utilizing the distance coherence which can be extracted from the GPS itself without the need to require other information. We applied three ensemble classification RandomForest, ExtraTrees, and Bagging algorithms; and the experimental result showed that the approach can achieve 99.42\%, 99.12\%, and 99.25\% of accuracy, respectively.

\end{abstract}

\section{Introduction}
\label{section:introduction}
The term Society 5.0~\cite{society50} has become a well-known buzzword which was introduced by the Japanese government in 2011\footnote{Society 5.0 follows Society 1.0 (the hunting society), Society 2.0 (agricultural society), Society 3.0 (industrial society), and Society 4.0 (information society}. Society 5.0 is used to refer to a super-smart society that balances economic advancement with the resolution of social problems. 
Society 5.0 focuses on two important keywords \textbf{human-centered} and \textbf{smart} life with the support of Artificial Intelligent (AI), Internet of Things (IoT), big data, and cutting-edge technologies. 

Let's consider an example of the electronic payment system. In 1871, Western Union debuted the electronic fund transfer (EFT) which allows people to send money to pay for goods and services without necessarily having to be physically present at the point-of-sale. In 1946, John Biggins was the inventor of the first bank-issued credit card which can be used to replace paper money (although the concept of using a card for purchases and the term credit card was described in 1887 by Edward Bellamy). In 2011, Google was the first company to launch a project of mobile wallet which can be used to replace physical cash and even credit cards. Nowadays, the cashless payment system has become a recent trend. Many digital wallet services appeared such as Apple Pay (from 2014), Google Pay (from 2015 as Android Pay and from 2018 as Google Pay), Rakuten Pay (from 2016), etc. 

The biggest challenge for such payment systems is how to authenticate (verify) the users. The current approach is to rely on the authentication of the mobile phones using PIN code, password, biometrics information (i.e., fingerprinting, iris, face, etc.), or multi-factor method which combines more than one method of authentication from independent categories of credentials.

\subsubsection{Attacks and Vulnerabilities in Current Smartphone Authentication}
Gradually, there appeared many sophisticated attacks in smartphone authentication. First, \emph{PIN code/ password-guessing attack}~\cite{hashcat, johnripper} tries to recover the password plaintext from its hashed form using brute force attack which systematically checks every combination of letters, symbols, numbers and dictionary attack which uses a dictionary of common words. 
Second, \emph{biometric spoofing} tries to generate synthetic or fake biometric traits of legal users to fool the capture sensors including \emph{facial spoofing} which utilizes printed facial photographs and digital video~\cite{facialvideo} or a 3D mask~\cite{facial3D}, \emph{fingerprinting spoofing}~\cite{fingerprintingsp} which utilizes artificial replicas with different materials such as gelatin, latex, play-doh or silicone, and \emph{iris spoofing}~\cite{iris} which utilizes an image forging natural iridal texture characteristics~\cite{irisoriginal} or even cosmetic contact lenses~\cite{contactlens, contactlens1}, and the combination of all these three spoofing types~\cite{combine}. Third, \emph{smudge attack} tries to guess the graphical password pattern in touch screen phones by analyzing the epidermal oils and smears left on the device's screen by the user's fingers~\cite{smudge}. Fourth, \emph{shoulder-surfing attack}~\cite{chi17} uses social engineering techniques to steal the victim's personal information such as PIN code and password by looking over the victim's shoulder or by eavesdropping sensitive information being spoken and heard or keystrokes on a device. 

Furthermore, not related to any attack, several studies found that a large number of users themselves do not lock their smartphones. An analysis of over 150 smartphone users was conducted in~\cite{introsoup2013} and showed that 33\% of the users do not use any screen lock even PIN, password, or pattern. Face-to-face qualitative interviews with 28 participants were conducted in~\cite{introccs2014}. 29\% of the users responded that they did not lock their devices. The three most common reasons include: emergency personnel not being able to identify them, not having the devices returned if lost, and not believing they worth data. An online survey with 260 participants and a field study with 52 participants was performed to analyze smartphone users' risk perception and behaviors~\cite{introsoup2014}. They showed that 40.9\% of users use slide-to-unlock and 16.2\% of users do not use any screen lock. 

\subsubsection{Location-based Behavioral Authentication}
Toward the construction of a smarter and securer mobile-based authentication system, there are several questions. First, for mitigating the aforementioned attacks, is there an additional mobile-based authentication method that can support the conventional methods such as using PIN code, password, and human biometric traits (i.e., fingerprints, face, iris)? Second, imaging the scenario that a person is on the way going to a coffee shop. Before the person arrives, the coffee shop can predict that he/she will arrive 15 minutes later with a high probability, and prepares in advance his/her usual order, and will automatically subtract the charge from his account. The person then does not need to wait time for the order and payment process. So, the question is: is it possible to authenticate and predict the location (for example, the coffee shop) that the users are likely going to? Last but not least, in the recent situation of the COVID-19 pandemic, the current smartphone-based cashless payment can reduce the chance of using cash or card, but still, the user needs to touch the smartphone screen to show the bar code to the cashier. The final question is that is it possible if the user can pay for goods when only bringing the smartphone without the need to touch the screen?

An idea that can answer these questions is using behavioral (or habit)-based information. It is a new research topic in which the main challenge is how to decide good behavioral information for authentication. Inspirited from L. Fridman (MIT) et al.~\cite{gps2016} just in 2016, GPS location history is the most promising approach because ``It is relatively unique to each individual even for people living in the same area of a city. Also, outside of occasional travel, it does not vary significantly from day to day. Human beings are creatures of habit, and in as much as location is a measure of habit''. At this time, it can only say that single behavioral authentication is an additional method to support the conventional methods (i.e., password, PIN code, biometrics) or a method to be combined with other behavioral factors. In the future, if a payment system can be constructed such that the users do not need to bring anything even small wearable devices such as smartwatches or RFID chips (e.g., the data can be collected via satellite sensors) and which can completely replace the conventional biometrics authentication, it is a step closer to Society 5.0. 


\subsubsection{Motivation}
A system that can achieve a high authentication accuracy is when it can collect multiple factors as much as possible. However, from the users' viewpoint, the most convenient system that does not bring strong privacy concerns to the users is when it does not require the users to provide too much information. From the GPS records, most of the previous work utilized the longitude, the latitude, and the information extracted from the timestamp (i.e., year, month, day, hour, minute, second, day of the week, etc.) as the features in the classification machine learning model for the user authentication. Given limited information from the GPS (longitude, latitude, and timestamp), if metadata that carries extra independent information can be obtained from the GPS itself, it can help to improve the accuracy. An example of GPS-based self-enhancement comes from~\cite{thaodbsec} in which the address is extracted from the pair of longitude and latitude using a reverse geocoding. 

\subsubsection{Contribution} 
In this paper, we propose an idea of extracting the distance coherence features from the GPS records themselves without the need to request any other information besides the GPS. For each user, the locations at close time clocks may have some closer correlation in physical distance than the locations at far time clocks since a human needs a period of time to gradually move from a location to another location. Since the idea actually reflects a movement ``lifestyle'' of the users, we hypothesized that it may improve the accuracy. Although it may be not 100\% correct when the user goes forward and then backward within the considered period of time, the proposed distance coherence features are used as the additional features to support the previous features. 

To evaluate how feasible the approach is, we collected 107,637 GPS records from 348 users. We applied three ensemble machine learning classification (RandomForest, ExtraTrees, and Bagging) on a total of 13 features including the distance coherences features. The experimental result showed that our approach outperforms the approach without the distance coherence features with the accuracy of 99.42\% (for RandomForest), 99.12\% (for ExtraTrees), 99.25\% (for Bagging) and merely 0\% of false positive rate and 0.01\% of false negative rate (for all the three algorithms).

Considering it reasonability, it may raise the discussion that since the distance coherence score can be inferred from the GPS and the timestamp, so whether the entropy of the distance coherence is the same as that of the GPS and the timestamp, or in other words, whether the distance coherence gives no additional information to the GPS and the timestamp. However, for each sample, the corresponding distance coherence is computed from not just the sample but also other samples that have a close timestamp with the considered sample. Therefore, the GPS, the timestamp, and the distance coherence score are independent variables. Furthermore, of course the model using GPS and timestamp can be improved if they are combined with other factors such as Wifi information, web browser log, etc. However, the goal in this paper is to make clear whether the distance coherence score extracted from the GPS and the timestamp can be helpful for the better classification model. We thus excluded other factors to make the comparison clean.

\subsubsection{Roadmap}
The rest of this paper is organized as follows. The related work is introduced in Section~\ref{section:realtedwork}. The proposed method is described in Section~\ref{section:proposedmethod}. The experiment is presented in Section~\ref{section:experiment}. The threat model is presented in Section~\ref{section:threatmodel}. The discussion about future work is shown in Section~\ref{section:future}. Finally, the conclusion is drawn in Section~\ref{section:conclusion}.

\section{Related Work}
\label{section:realtedwork}

In this section, we present related work focusing on multimodal authentication using human-smartphone interactions and using other factors. The term \emph{multimodal} (not \emph{multimodel}) is used to indicate the biometrics authentication using multiple biometric data. It is opposite with \emph{unimodal} which uses only a single biometric data. 

\subsection{Multimodal Authentication for Smartphone}
L. Fridman et al.~\cite{gps2016} analyzed the behavioral data of four modalities from active mobile devices including text stylometry typed on a soft keyboard, application usage patterns, web browsing behavior, and physical location of the device from GPS (outdoor) and Wifi (indoor). The data was collected from 200 users in more than 30 days. The authors proposed  a parallel binary decision-level fusion architecture for classifiers based on four biometric modalities. A. Alejandro et al.~\cite{gpsMULEA2019} performed an analysis on a multimodal data consisting of four biometric data channels (including touch gestures, keystroking, accelerometer, and gyroscope) and three behavior profiling (including WiFi, GPS location and app usage). The data was obtained during the natural human-smartphone interaction of 48 users on average 10 days per user. They proposed two authentication models named the one-time approach that uses all the channel information available during one session, and an active approach that uses behavioral data from multiple sessions by updating a confidence score. W. Shi et al.~\cite{gps2011} proposed an authentication framework that enables continuous and implicit user identification service for a smartphone. The data was collected from four sensor modalities including voice, GPS location, multitouch, and locomotion. They conducted a preliminary empirical study with a small set of users (seven). The result showed that the four modalities are enough for mobile user identification. R. Valentin et al.~\cite{gpsacm2018} conducted an analysis on multimodal sensing modalities with mobile devices when the GPS, accelerometer, and audio signals are utilized for human recognition. The data was collected from four existing datasets which consist of 491 users. They applied four variants of deep learning for interpreting user activity and context as captured by multi-sensor systems. M. Upal et al.~\cite{gpsupal2016} investigated user authentication methods using the first non-commercial multimodal data which focuses on three smartphone sensors (front camera, touch sensor, and location service). The data was collected from 48 users for 2 months. Their benchmark results for face detection, face verification, touch-based user identification, and location-based next place prediction showed that more robust methods fine-tuned to the mobile platform are needed to achieve satisfactory verification accuracy.  T. Thao et al.~\cite{thaodbsec} extracted the addresses given the longitudes and latitudes from the GPS records and then applied the text mining on the addresses. The data was collected from 50 users for about four months. Their experimental result showed that the combination between the text features and the GPS data can improve the classification accuracy. B. Aaron et al.~\cite{gpsuspatent2018} proposed a wallet repository that can store biometric data using multiple layers: biometric layer, a genomic layer, a health layer, a privacy layer, and a processing layer. The processing layer can be used to determine and track the user location, the speed when the user is moving using GPS data. 

\subsection{Other Multimodal Authentication}

Besides using human-smartphone interactions, multimodal authentication also uses other factors. T. Kaczmarek et al.~\cite{other1} investigated a new hybrid biometric based on a human user’s seated posture pattern in an average office chair over the course of a typical workday. Their experimental results on a population of 30 users showed that the posture pattern biometric can capture a unique combination of physiological and behavioral traits and can authenticate the users with 91\% of accuracy. M. Ivan et al.~\cite{other2} proposed an approach which combines the PIN code and the pulse-response. For the experiment process, the biometric information from 10 users was collected. The result showed that each human body exhibits a unique response to a signal pulse applied at the palm of one hand, and measured at the palm of the other. The experimental result for user authentication achieved 88\% of accuracy when the records are taken weeks apart. W. Louis et al.~\cite{signal1} and R. Alejandro et al.~\cite{signal2} constructed a continuous authentication system based on electrocardiogram (ECG) and electroencephalogram (EEG). Their approaches achieved 1.57\% and 0.82\% of false negative rate, respectively. E. Simon et al.~\cite{othereyemovement} extracted distinct patterns from eye movement (it is different from iris) with 21 features that can be used for user authentication. The data was collected from 30 users in 2 weeks with 3 scenarios (no prior knowledge, the knowledge gained through a description, and  knowledge gain through observation). The experimental result achieved 3.98\% of equal error rate. 

\section{Proposed Approach}
\label{section:proposedmethod}

In this section, we describe our proposed method including data collection, feature extraction and selection, and our learning method. 

\subsection{Data Collection}
\label{section:datacollection}
We created a navigation application named MITHRA (Multi-factor Identification/auTHentication ReseArch) in the project of the University of Tokyo to collect the GPS information of the users. The application can be installed on both iOS and Android smartphones. The application was developed to be run in the background so that the users do not feel a burden with the user interface (UI) and that the memory consumption can be saved. The data was collected from 348 users with 107,637 GPS records including pairs of longitude and latitude for four months from January 11th to April 26th in 2017\footnote{Although the GPS was collected from smartphones in this project, GPS can completely be collected from many smaller devices such as smartwatch or smartband nowadays}. Compared to the existing works (see Secion~\ref{section:realtedwork}), the number of users in our dataset is higher than most of the papers and is only lower than~\cite{gpsacm2018} which could collect the information from 491 users. We recruited the participants randomly. The users live and work in random areas. The GPS data was measured every minute. The value of the longitudes and latitudes were collected with the precision up to 6 decimal places (e.g., 36.xxxxxx) corresponding to 0.1 meters. 

\subsubsection{Privacy Consent}
The privacy consent is shown to the users during the installation process. The installation can only be done if the users accept the terms and conditions agreement. Even after the application is successfully installed, the users can choose to start or stop using the application anytime. Any personal information of the users such as name, age, gender, race, ethnicity, income, education, etc. is not collected. Only the email address is collected as the user identity in the collected data which is used to distinguish the users with each other. Even though the application is used to collect the GPS information, the users do not need to disclose which location is the home, which location is the office, etc. Our project was reviewed by the Ethics Review Committee of the Graduate School of Information Science and Technology, the University of Tokyo. Finally, all the users who installed the application agreed to participate in our project. 

\subsection{Feature Extraction and Selection}
The features are categorized into two groups: (i) the features extracted from the GPS and the timestamp, and (ii) the features using the distance coherence score.  

\subsubsection{GPS and Datetime}
There are seven features in this group. Two features were extracted from the GPS including the latitudes and the longitudes which are represented by float numbers. The valid ranges for the latitudes and the longitudes are the continuous range $[-90, + 90]$ and $[-180, +180]$, respectively.  Five features were extracted from the timestamp including month, day, hour, minute, and day of a week (i.e., seven days from Monday to Sunday) which are represented by integer numbers. The valid ranges for these features are the intervals $[1, 12]$, $[1, 31]$, $[0, 23]$, $[0, 59]$, and $[1, 7]$, respectively. The year was not extracted as a feature because all the samples in the dataset were collected in the same year (2017). 

\subsubsection{Distance Coherence}
\label{section:featuredc}
There are $\alpha$ features in this group (we will soon explain how to choose $\alpha$). Every $z$-th feature (where $z \in [1, \alpha]$) represents the distance coherence (also known as similarity score) between each sample in the dataset with the average of all the other samples in the dataset that belong to the same user and that occur before or after $p$ hours for every $p \in [0, z]$ with the considered sample. $p=0$ is the case when the other samples occur in the same hour with the considered sample. 

\begin{figure}[!ht]
\centering
\includegraphics[scale=0.3]{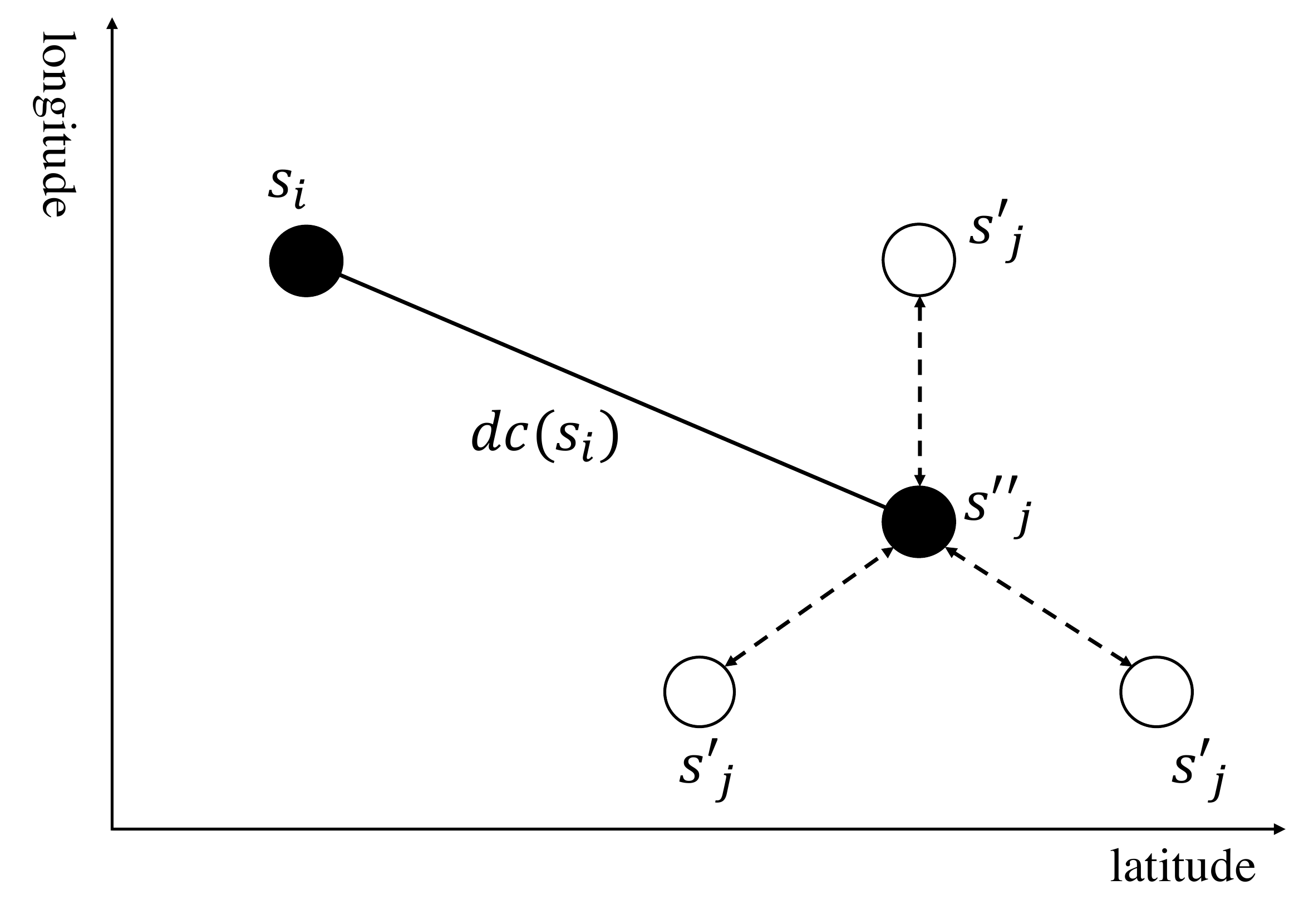}
\caption{Distance Coherence (Similarity Score)}
\label{fig:distance}
\end{figure}

More concretely, the features are computed as follows (see Figure~\ref{fig:distance}). Let $\{\mathsf{dc}_{z}\}$ denote the set of $\alpha$ features where $z \in [1, \alpha]$. Let $s_{i}$ denote each sample in the dataset where $i \in [1, n]$ and $n$ denotes the number of samples (in our dataset, $n = 107,637$). For each feature $\mathsf{dc}_{z}$, let $K_{z} = \{s'_{j}\}$ (where $j \in [1, n]$ and $j \neq i$) denotes the set of all the other samples such that $s_{i}$ and $s'_{j}$ belong to the same user $U_{t}$ (where $t \in [1, 348]$). State differently,  $s_{i}$ and $s'_{j}$ have the same label $U_{t}$. Let $lat(s_{i})$ and $lat(s'_{j})$, $lon(s_{i})$ and $lon(s'_{j})$, and $hour(s_{i})$ and $hour(s'_{j})$ denote the latitude, the longitude, and the hour features for $s_{i}$ and $s'_{j}$, respectively. For each $\mathsf{dc}_{z}$, $K_{z}$ is chosen such that: 


\begin{equation}
\label{eq:hour}
hour(s_{i}) = hour(s'_{j}) \pm p \quad\quad \text{for } \forall p \in [1, z] 
\end{equation}
The average coordinate $s''_{j}$ is determined from all the samples $s'_{j}$ in $K_{z}$ such as:
\begin{equation}
lat(s''_{j}) = \mathsf{average}(lat(s'_{j})) \quad \forall s'_{j} \in K_{z} 
\end{equation}
\begin{equation}
lon(s''_{j}) = \mathsf{average}(lon(s'_{j})) \quad \forall s'_{j} \in K_{z}
\end{equation}
The features are finally calculated as the distance between $s_{i}$ and $s''_{j}$:

\begin{equation}
\label{eq:dc1}
\mathsf{dc}_{z}(s_{i}) = \sqrt[2]{(lat(s_{i})- lat(s''_{j}))^{2} + (lon(s_{i})- lon(s''_{j}))^{2}} 
\end{equation}

From Equation~\ref{eq:hour}, we can observe that $K_{z}$ chosen for $\mathsf{dc}_{z}$ is a subset of $K_{z'}$ chosen for $\mathsf{dc}_{z'}$ for all $z, z' \in [1, \alpha]$ such that $z'>z$. It may raise the question that whether all the $\alpha$ features have a correlation. However, the averages from even correlated sets are completely different (for example, $\mathsf{average}(1, 2, 3) = 2$ which is different from $\mathsf{average}(1, 2, 3, 4) = 5$). All the features $\mathsf{dc}_{z}$ are thus independent variables. A numeric example for how to calculate the distance coherence features will be given in Appendix~\ref{section:appendixexample}.

We now explain how the concrete value for $\alpha$ is. In our approach, we use three advanced classification machine learning algorithms which are RandomForest, ExtraTrees, and Bagging (explained in more detail in Section~\ref{section:learning}). We conducted an experiment for every $\alpha$ from 1 and increased it gradually. We found that the best $\alpha$ for RandomForest, ExtraTrees, and Bagging is 3, 4, and 5, respectively, at which the algorithms reach a peak performance (Section~\ref{section:bestalpha}). Since $\alpha$ reflects the movement lifestyle of the users, it is reasonable for $\alpha$ to be not large. For instance, the GPS record (a pair of latitude, longitude) of a user $U_{t}$ at 15:00 may have some correlation (in terms of physical distance) with that at 14:00 and 16:00 than that at 13:00 and 17:00 or than that at 12:00 and 18:00. In the rest of this paper, we use $\alpha$-DC to denote the approach in which $\alpha$ distance coherence features are used, and $\{\mathsf{lat}$, $\mathsf{lon}$, $\mathsf{mon}$, $\mathsf{day}$, $\mathsf{hour}$, $\mathsf{min}$, $\mathsf{weekday}$, $\mathsf{dc}_{1}$, $\mathsf{dc}_{2}$, $\cdots$, $\mathsf{dc}_{6}\}$ to denote the set of the thirteen features related to both the GPS and timestamp and the distance coherence.

\subsubsection{Feature Distribution}
The distribution statistics for the features is described in Table~\ref{table:distribution} which includes the mean, standard error, median, standard deviation, Kurtosis score, skewness score, min value, and max value. A normal distribution check for the features is not necessary~\cite{ACHI}. 
The negative and positive values in the latitude and the longitude in the ``Min'' and ``Max'' columns indicate that the users who used to commute in Japan might travel abroad during the time of data collection. This kind of data can create noises during the training and testing processes. However, we do not think it should be removed because the data reflects the natural behavior of the users. Although the noises may lower the accuracy, we want to measure how practical the approach is when using real data without being manipulated. 

\begin{table*}[!ht]
\centering
\caption{Feature Distribution}
\label{table:distribution}
\footnotesize
\begin{tabular}{l r r r r r r r r}
\textbf{Feature} & \textbf{Mean} & \textbf{SE} & \textbf{Median} & \textbf{SD} & \textbf{Kurtosis} & \textbf{Skewness} & \textbf{Min} & \textbf{Max} \\
\hline
$\mathsf{lat}$ & 35.262 & 0.014 & 35.376 & 4.554 & 151.722 & $-$10.935 & $-$36.858 & 43.907\\
$\mathsf{lon}$ & 136.783 & 0.034 &  137.846 & 11.165 & 248.09 & $-$15.101 & $-$121.979 & 174.799 \\
$\mathsf{month}$ & 3.321 & 0.002 & 3.000 & 0.753 & $-$0.260 & $-$0.777 & 1.000 & 4.000 \\
$\mathsf{day}$ & 17.328 & 0.026 & 19.000 & 8.600 &  $-$1.075 & $-$0.285 & 1.000 & 31.000 \\
$\mathsf{hour}$ & 13.421 & 0.019 & 14.000 & 6.388 & $-$0.820 & $-$0.417 & 0.000 & 23.000\\
$\mathsf{min}$ & 28.919 & 0.053 & 29.000 & 17.357 & $-$1.186 & 0.038 & 0.000 & 59.000\\
$\mathsf{weekday}$ & 3.986 & 0.006 & 4.000 & 1.966 & $-$1.215 & $-$0.016 & 1.000 & 7.000 \\
$\mathsf{dc}_{1}$ & 4,104.198 & 137.84 & 191.318 & 45,214.683 & 976.703 & 27.756 & 0.000 & 2,545,473.711\\
$\mathsf{dc}_{2}$ & 4,359.489 & 137.598 & 239.163 & 45,140.323 & 988.797 & 27.801 & 0.000 & 2,548,301.562 \\
$\mathsf{dc}_{3}$ &  4,586.805 & 140.910 &  259.640 & 46,228.488 & 995.124 & 27.895 & 0.000 & 2,549,190.471 \\ 
$\mathsf{dc}_{4}$ & 4,678.671 & 140.658 & 272.654 & 46,147.07 & 978.139 & 27.653 & 0.004 & 2,554,832.383 \\
$\mathsf{dc}_{5}$ & 4,781.704 & 141.784 & 276.978 & 46,516.486 & 1,002.699 & 28.001 & 0.048 & 2,567,773.385 \\
$\mathsf{dc}_{6}$ & 4,822.694 & 143.361 & 284.604 & 47,033.864 & 1,013.685 & 28.284 & 0.017 & 2,568,234.888 \\
\end{tabular}\\
$ $\\
SE (Standard Error), SD (Standard Deviation), DC: Distance Coherence
\end{table*}

\subsection{Learning}
\label{section:learning}

This section explains the machine learning algorithms chosen for our model and the evaluation method. In the dataset, each user has a different label. Each label has a different set of records.

\subsubsection{Average Ensemble Classifications}

The dataset contains 107,637 samples with a large number of labels (348 users). Instead of using the traditional algorithms, we use the advanced algorithms called \emph{average ensemble classifications} to get better performance in terms of training speed and accuracy. The average ensemble technique builds several base estimators independently and produces one optimal predictive estimator by averaging the predictions of all the base estimators. The combined estimator is better than any of the single base estimators by reducing the variance to control over-fitting. The most common average ensemble algorithms are the followings:

\begin{itemize}

\item RandomForest~\cite{RandomForest2001}: This algorithm implements a meta estimator that fits a number of decision tree classifiers on various randomized sub-samples of the dataset and uses averaging to create the best predictive estimator. When each estimator is built, a bootstrap is created by randomly sampling the dataset with replacement. The size of the sub-samples is set to be the same as the size of the original input sample. A decision tree is usually trained by recursively splitting the data (the process to convert the non-homogeneous parent into the two most homogeneous child nodes). The algorithm selects an optimal split on the features selected at every node 

\item ExtraTrees~\cite{ExtraTrees2006}: This algorithm also produces the best predictive estimator in a way like RandomForest. However, there are some differences. First, while RandomForest uses the optimal split, ExtraTrees uses the random split. Second, while RandomForest sets the $bootstrap=True$ by default, ExtraTrees sets the $bootstrap=False$ by default. This also means that while RandomForest supports drawing sampling with replacement, ExtraTrees supports drawing sampling without replacement. 

\item Bagging (Bootstrap Aggregating)~\cite{Bagging2012}: This algorithm has one different point from RandomForest and ExtraTrees. While RandomForest and ExtraTrees select only a subset of randomized features for splitting a node, Bagging uses all the features for splitting a node.
\end{itemize}

\subsubsection{Stratified K-Fold}
\begin{figure}[!ht]
\centering
  \includegraphics[scale=0.33]{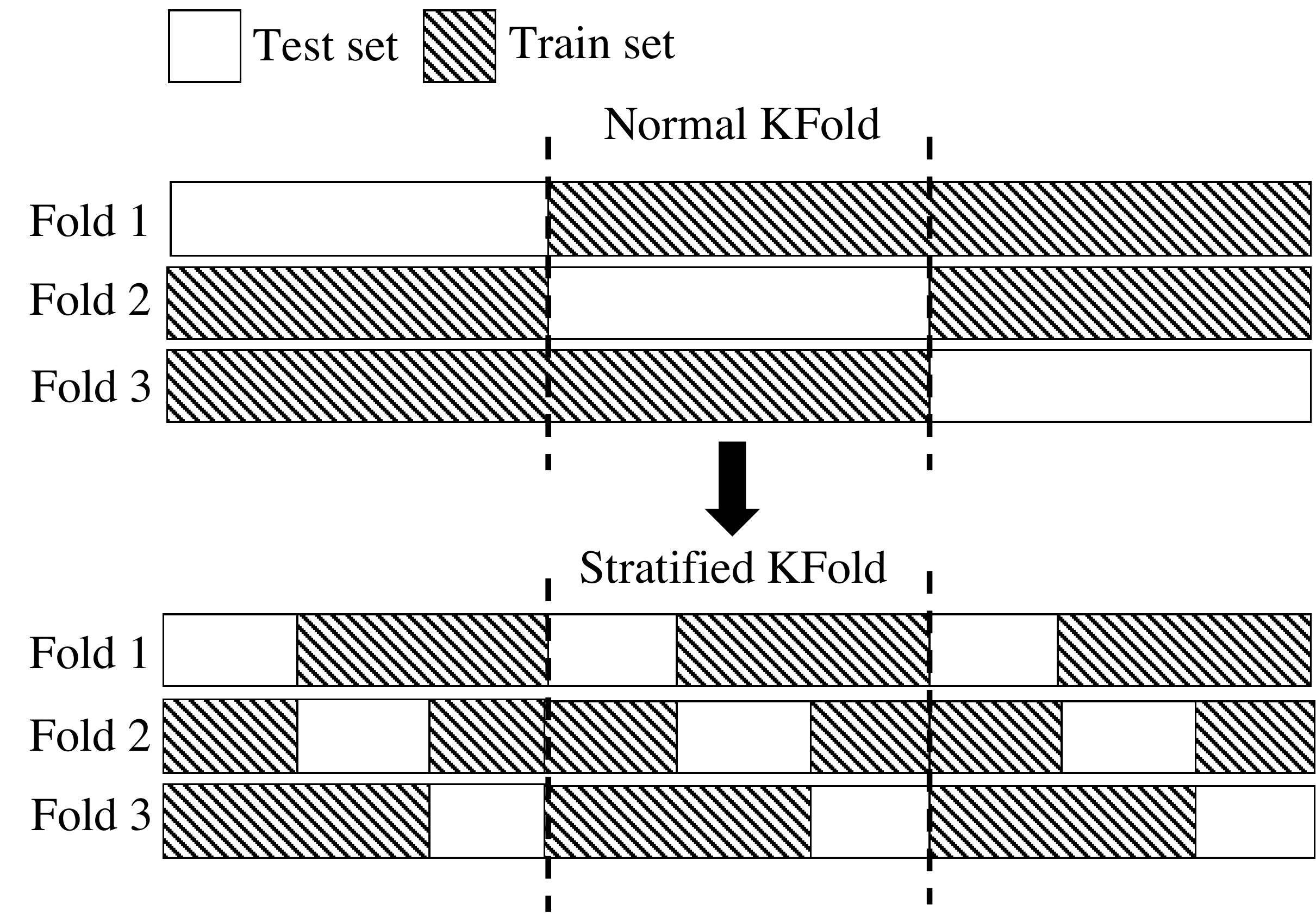}
  \caption{A Stratified KFold}
  \label{fig:kfold}
\end{figure}

The data is shuffled at first and is then used in a $k$-fold cross validation. Since the numbers of samples of the users are imbalanced, using the normal $k$-fold cross validation can lead to the following problem. There may exist a class $c_{k}$ where $k \in \{1, 2, \cdots, 348\}$ such that all the samples in its sample set $S_k$ belong to the test set, and the training set does not contain any of its samples. The classifier, therefore, cannot learn about the class $c_{k}$. To solve this problem, we used \emph{Stratified $k$-fold} cross-validation object which is a variation of $k$-fold and can deal with imbalanced data in each class. As presented in Figure~\ref{fig:kfold}, it splits the data in the train and the test sets and returns stratified folds that are made by preserving the percentage of samples for each class. 

\subsubsection{Evaluation Metrics} To evaluate our approach, we measure the following metrics: 
\begin{equation}
accuracy = \frac{tp+tn}{tp+fp+fn+tn}, \quad precision = \frac{tp}{tp+fp}, \quad recall = \frac{tp}{tp+fn}
\end{equation}
\begin{equation}
F1 = 2 \times \frac{recall \times precision}{recall+precision}, \quad FPR = \frac{fp}{fp+tn}, \quad  FNR = \frac{fn}{fn+tp}
\end{equation}
where $tp, tn, fp, fn$ denote the true positive, true negative, false positive, and false negative values, respectively. $FPR$ and $FNR$ denote the false positive rate and false negative rate, respectively. The accuracy is a good metric when the distribution for each label is almost similar. However, for an imbalanced dataset, F1-score is the better metric. 

\section{Experiment}
\label{section:experiment}

This section presents the experimental setup, the results obtained after applying the classification, and how to find the best $\alpha$ for each algorithm.

\subsection{Experimental Setup}
The program was implemented using Python 3.7.4 on a computer MacBook Pro 2.8 GHz Intel Core i7, RAM 16 GB. The machine learning algorithms are executed using \emph{scikit-learn}\footnote{https://scikit-learn.org/stable/} library version 0.22.

For each ensemble algorithm, the number of base estimators $n\_estimators$ is set to 100. The $k$ value in the stratified $k$-fold cross validation is set to $k = 10$. Since the categorical labels are represented in text strings (such as ‘user001’, ‘user002’, etc.), the labels are transformed to numerical values using the \emph{label encoding}. Contrary to the \emph{ordinal encoding} which encodes the labels to an integer array and the \emph{one-hot encoding} which encodes the labels to a one-hot numeric array, the label encoding encodes the labels to the values between 0 and $q-1$ where $q$ is the number of distinct labels of all the classes. The label encoding is the most lightweight method and uses less disk space. Since the data is imbalanced, to avoid the situation that F1 is not between precision and recall, we calculate the three metrics (precision, recall, and F1 score) for each label and find their average weight by the number of true instances of each class. This process can be done by setting the parameter $average = weighted$ in the $sklearn.metrics$. For the accuracy, this parameter is not necessary. Since the values of the distance coherence features are small, we scaled them up $\times10^{4}$.  For each of the three algorithms (RandomForest, ExtraTrees, and Bagging), an experiment was conducted with different $\alpha$'s . The classification was applied on 107,637 samples with 348 labels which correspond to 348 users.

\subsection{Main Result}

The main result is presented in Table~\ref{table:main}. In the table, NoDC represents the approach not using distance coherence features while $\alpha$-DC represents the approach using $\alpha$ distance coherence features. As proved later in Section~\ref{section:bestalpha}, RandomForest, ExtraTrees, and Bagging reach the best performance at $\alpha = 3$, $\alpha = 4$ and $\alpha = 5$, respectively. Thus, 3-DC, 4-DC, and 5-DC are chosen to compare with NoDC in this table (although only 1-DC can already beat NoDC (see Section~\ref{section:bestalpha})). 

The result shows that our approach $\alpha$-DC outperforms NoDC in all the cases. Comparing all the algorithms using NoDC only with each other, Bagging gives the best result with 98.69\% of F1 score with 0.02\% of false negative rate. Comparing all the algorithms using our approach with each other, RandomForest gives the best result with 99.42\% of F1 score and merely 0.01\% of false negative rate even though RandomForest just reaches $\alpha$ at $\alpha = 3$ (which is less than $\alpha = 4$ for ExtraTrees and $\alpha = 5$ for Bagging). Comparing the improvement between $\alpha$-DC and NoDC, ExtraTrees gives the best result when 2.34\% of F1 score is increased ($\triangle = +2.34$) and 0.04\% of false negative rate is reduced ($\triangle = -0.04$). 

\begin{table*}[!ht]
\centering
\caption{Result for Distance Coherent with Different Ensemble Algorithms\\$ $}
\label{table:main}
\begin{tabular}{l | r  r  r | r r r | r r r }
\multirow{2}{*}{\textbf{Measure }}  & \multicolumn{3}{c|}{\textbf{RandomForest}}  &  \multicolumn{3}{c|}{\textbf{ExtraTrees}}  &  \multicolumn{3}{c}{\textbf{Bagging}}  \\
& \textbf{ NoDC} & \textbf{3-DC} &  $\bm{\triangle} $ & \textbf{NoDC} & \textbf{4-DC}  & $\bm{\triangle} $ & \textbf{NoDC} & \textbf{5-DC} & $\bm{\triangle} $\\
\hline
F1 & 97.95 &  99.42 & \color{red} $\bm+$\textbf{1.47} & 96.77 &  99.11 & \color{red}  $\bm+$\textbf{2.34} & 98.69 & 99.24  & \color{red} $\bm+$\textbf{0.55} \\
Accuracy & 97.97 & 99.42  &  \color{red}  $\bm+$\textbf{1.45} &  96.80 &  99.12& \color{red}  $\bm+$\textbf{2.32} & 98.69 &  99.25 & \color{red} $\bm+$\textbf{0.56}\\
Precision & 98.05 & 99.45  & \color{red} $\bm+$\textbf{1.40} & 96.90 &  99.15 & \color{red} $\bm+$\textbf{2.25}  & 98.75 &99.28  & \color{red} $\bm+$\textbf{0.53} \\
Recall & 97.97 & 99.42  & \color{red}  $\bm+$\textbf{1.45} & 96.80 & 99.12 & \color{red} $\bm+$\textbf{2.32}  & 98.69 & 99.25 & \color{red} $\bm+$\textbf{0.56} \\
FPR & 0.00& 0.00 & \color{red}  \textbf{0.00} & 0.00&  0.00 & \color{red}  \textbf{0.00} & 0.00 &  0.00 &  \color{red} \textbf{0.00} \\
FNR & 0.03& 0.01 & \color{red} $\bm-$\textbf{0.02} & 0.05 & 0.01 & \color{red} $\bm-$\textbf{0.04} & 0.02 & 0.01 & \color{red} $\bm-$\textbf{0.01} \\
\end{tabular}
$ $\\
NoDC: the approach without distance coherence features, \\$\alpha$-DC ($\alpha = 3, 4, 5$): the approach using distance coherence features, \\$\triangle$: the improved score between $\alpha$-DC and NoDC.
\end{table*}

\subsection{Best Alpha ($\bm{\alpha}$) for Each Algorithm}
\label{section:bestalpha}

\begin{figure}[!ht]
\centering
\includegraphics[scale=0.48]{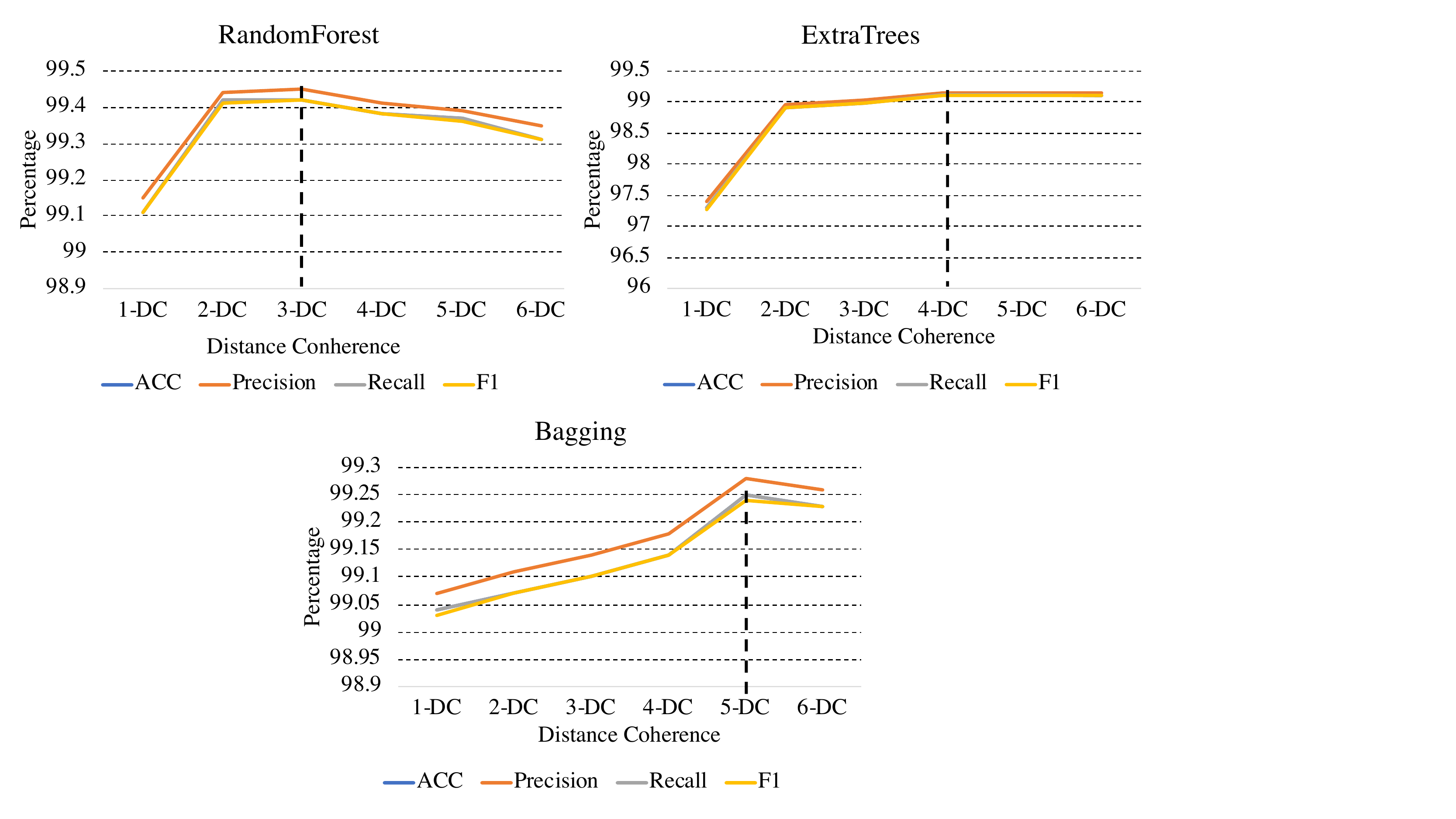}
\caption{Different Alpha's for Distance Coherence}
\label{fig:alpha}
\end{figure}

This section explains the experiment to find the best $\alpha$ for each algorithm. First, $\alpha$ is set to 1 and is then gradually increased until the performance becomes convergent or is reduced after reaching the peak. The result and its graphs are presented in Table~\ref{table:alpha} and Figure~\ref{fig:alpha}. The proposed approach using RandomForest, ExtraTrees, and Bagging got the best performance at $\alpha = 3$, $\alpha = 4$, and $\alpha = 5$, respectively. Figure~\ref{fig:alpha} shows that in all the algorithms, the graph almost has the cone shape (the result is gradually increased, gets the peak, and then is reduced or becomes convergent), not a zigzag shape (in which we cannot predict where is the peak). The result also shows that by even just using 1-DC ($\alpha = 1$), our approach can already beat NoDC.

\begin{table*}[!ht]
\centering
\caption{Result for Each Alpha\\$ $}
\label{table:alpha}
\begin{tabular}{l l r r r r r r}
& & \textbf{1-DC} & \textbf{2-DC} & \textbf{3-DC} & \textbf{4-DC} & \textbf{5-DC} & \textbf{6-DC} \\
\hline
\multirow{6}{*}{\textbf{RandomForest}} & F1 &  99.11 & 99.41 & \color{red} \textbf{99.42} & 99.38 & 99.36 & 99.31\\
& Accuracy & 99.11 & 99.42 & \color{red}\textbf{99.42} & 99.38 & 99.37 & 99.31\\
& Precision & 99.15 & 99.44 & \color{red}\textbf{99.45} & 99.41 & 99.39 & 99.35\\
& Recall & 99.11 & 99.42 & \color{red}\textbf{99.42} & 99.38 & 99.37 & 99.31\\
& FPR & 0.00 & 0.00 & \color{red}\textbf{0.00} & 0.00 & 0.00 & 0.00\\
& FNR & 0.01 & 0.01 & \color{red}\textbf{0.01} & 0.01 & 0.01 & 0.01\\
\hline
\multirow{6}{*}{\textbf{ExtraTrees}} & F1 & 97.27 & 98.90 & 98.98 & \color{red} \textbf{99.11} & 99.11 & 99.11\\
& Accuracy & 97.30 & 98.91 & 98.99 & \color{red}\textbf{99.12} & 99.12 & 99.11\\
& Precision & 97.40 & 98.95 & 99.03 & \color{red}\textbf{99.15} & 99.15 & 99.15\\
& Recall & 97.30 & 98.91 & 98.99 & \color{red}\textbf{99.12} & 99.12 & 99.11\\
& FPR & 0.00 & 0.00 & 0.00 & \color{red}\textbf{0.00} & 0.00 & 0.00\\
& FNR & 0.04 & 0.02 & 0.02 & \color{red}\textbf{0.01} & 0.01 & 0.01\\
\hline
\multirow{6}{*}{\textbf{Bagging}} & F1 &  99.03 & 99.07 & 99.10 & 99.14 &\color{red} \textbf{99.24} & 99.23\\
& Accuracy & 99.04 & 99.07 & 99.10. & 99.14 & \color{red}\textbf{99.25} & 99.23\\
& Precision & 99.07 & 99.11& 99.14 & 99.18 & \color{red}\textbf{99.28} & 99.26\\
& Recall & 99.04 & 99.07 & 99.10 & 99.14 &\color{red} \textbf{99.25} & 99.23\\
& FPR & 0.00 & 0.00& 0.00 & 0.00 & \color{red}\textbf{0.00} & 0.00\\
& FNR & 0.01 & 0.01 & 0.01 & 0.01 & \color{red} \textbf{0.01} & 0.01\\
\end{tabular}
\end{table*}

\subsection{Computation Time}
For the best algorithms (5-DC using Bagging, 4-DC using ExtraTrees, and 3-DC using RandomForest), the average computational time for the training and cross validation processes from 5 execution times is 2,272 seconds (38 minutes), merely 270 seconds (4.5 minutes), and 596 seconds (10 minutes) respectively. It is not a big deal for the server. When the number of users is much more increased (e.g., to thousands), it is not complicated to transform the current model from the one-class classification to a multi-class classification where each user has a different classifier with binary labels representing whether or not a sample belongs to that user.

\section{Threat Model}
\label{section:threatmodel}

In this section, we present the threat model including which attack is focused on, the adversary's probability, and the assumptions.

\subsection{Targeted Attack}
Most of such authentication systems, not just our approach but other previous biometrics-based authentication, focus on protecting against insider threats in which the adversary tries to impersonate the authentication of an authorized user in the system. As mentioned in Section~\ref{section:introduction}, at this time the behavioral-based authentication should be used as an additional approach to support the conventional PIN code, password, or biometric authentications. So let's run an example in which our approach is combined with PIN code-based authentication. Let $Pr_{\mathcal{A}}$ denote the probability that the adversary $\mathcal{A}$ can break the system. $Pr_{\mathcal{A}}$ is defined as:
\begin{equation}
Pr_{\mathcal{A}} = Pr_{guess} \cdot Pr_{forge}
\end{equation}
where $Pr_{guess}$ and $Pr_{forge}$ denote the probability that $\mathcal{A}$ can correctly
guess the PIN code and the average probability that $\mathcal{A}$ can fool the classifier, respectively. $Pr_{forge}$ is the false negative rate which is the percentage of identification instances in which the unauthorized users are incorrectly accepted. Table~\ref{table:main} shows that all the 3-DC, 4-DC, and 5-DC approaches corresponding to the three different algorithms have the same 0.01\% of false negative rate. Thus, $Pr_{forge} = 10^{-4}$. Let $\tau$ and $\sigma$ denote the number of digits in the PIN code and the number of guessing candidates for each PIN code digit. If $\mathcal{A}$ has $n_{t}$ tries before the device is locked with many wrong PIN codes, we have $Pr_{guess} = \frac{n_{t}}{\sigma^{\tau}}$. Finally, $Pr_{\mathcal{A}}$ is thus:

\begin{equation}
Pr_{\mathcal{A}} = 10^{-4} \cdot \frac{n_{t}}{\sigma^{\tau}}
\end{equation}
Most of the new smartphone operation systems nowadays require 6 digits for PIN code. Typically, there are 10 digits of candidates from 0 to 9 for each digit. The users often have 4 to 6 PIN code tries for both Android and iOS before the device is locked. Therefore, $Pr_{\mathcal{A}} \simeq 4 \cdot {10^{-10}}$ to $6 \cdot {10^{-10}}$. 

Suppose the attacker can guess the PIN code after shoulder surfing and then robs the smartphone of the user. Since the application is designed such that every GPS record is sent to the server in realtime and the GPS history is not stored in the user smartphone, the attacker cannot see the log from the robbed phone to imitate the user's behavior. Also, there is no function of downloading the GPS log from the server to the smartphone because it is a doubtable action from a (suspicious) user. The only action that the attacker can manipulate on the GPS tracking application is to turn it on/off or uninstall it. If the attacker continues to use the smartphone (without the ability to search for the history log from the smartphone application), the probability for the attacker $Pr_{\mathcal{A}}$ is now 0.01\%. Even though it is not 0\% for the best case, it is still much better than 100\% for $\mathcal{A}$ to break the system without our approach. Similarly, if the \emph{collusion attack} in which an authorized user shares his/her PIN code to others occurs,  $Pr_{\mathcal{A}}$ is also 0.01\%. If the colluded user tells others his/her personal location history, it is unlikely for every single continuous GPS record to be imitated. This is why the idea of using behaviors (especially, long-term and continuous) is investigated. 

The model is assumed that the server storing the GPS cannot be accessed or corrupted by the adversary. The data is encrypted and only the trusted server can decrypt it. The data is transmitted via a secure network. Each smartphone is used by only a unique user. The smartphone and the server are protected against the side-channel attack which can collect the user data via timing information, power consumption, electromagnetic leaks, or sound. Last but not least, the users are assumed to be honest in sending their own data to the server where the classifier is performed because the data may be actively manipulated by an adversary seeking to make the classifier produce false negatives~\cite{adversarialclassifier}.

\subsection{Security Scenario Discussion}
In this section, we discuss other security scenarios from using smartphones.

\subsubsection{What if two users live and work in the same areas?}
As mentioned in Section~\ref{section:datacollection}, since our project recruited the users randomly, the users live and work in random areas. Even if there is a very rare case when two users live and work in the same area, they cannot have the same GPS tracking for every single hour because each user has many different activities at different timestamp not just at home and office (such as shopping, outdoor exercising, picking children at schools, etc.). Furthermore, inside the home and the office building, indoor positioning can be collected besides the GPS such as WiFi or Bluetooth beacons. Since the goal in this paper is to investigate the benefit of the extra information (i.e., the distance coherence) from the GPS itself, we do not consider to collect indoor location information; however, it is completely possible since the GPS and the indoor location information can be collected independently. Let's consider the case when legal users have the same trajectory within a period of time (e.g., elderly people in a senior home have daily activities confined to the surroundings). Since the longitude and latitude values have 6 decimal places (see Section~\ref{section:datacollection}), the precision is 0.1 meters. With this precision, two users cannot have the same movement log in a long period.

\subsubsection{How does the system work when individuals are outside their routine or when the attacker follows (imitates) the user’s behavior?}
Since these questions are not just for the GPS-based location authentication but the general behavioral-based authentication, we discuss from the general to specific perspectives. We want to emphasize that a single-factor behavioral-based authentication is used to support (not to replace) the conventional approaches such as password or biometrics; or it is combined with other behavioral factors to build up a multi-factor behavioral-based authentication. If a user is outside his/her routine or the attacker tries to imitate the user's behavior, the password/biometric or other routines are used to lower the false rejection and false acceptance rates. Although behavioral-based authentication has not yet been commonly used, this new but promising research has been proved to be possible for real applications. For instance, Google has launched the Project Abacus~\cite{Google1} just since 2016 to collect smartphone sensor signals (i.e., front-facing camera, touchscreen and keyboard, gyroscope, accelerometer, magnetometer, ambient light sensor, etc) and demonstrated that human kinematics can convey important information about user identity and can serve as a valuable component of multi-modal authentication systems. Among many behaviors, location is a typical factor to identify users. Human beings are creatures of habit, and in as much as location is a measure of habit~\cite{gps2016}. Also, the location is easy to collect since it is available in most modern smartphones. 

\subsubsection{Is it a problem when a user gets a new phone?}
It has no problem since the smartphone is just the device/tool, not the method. The user can register a location-based authentication system with an account and its application installed in his smartphones. As long as the user does not share his account to others and as long as the application is designed such that at a specific timestamp, an account can only be logged in a smartphone, his unique GPS data can be collected regardless of how many smartphones are used and regardless of whether the user shares his smartphones to others. 

\section{Future Work}
\label{section:future}

\begin{figure}[!ht]
\centering
\includegraphics[scale=0.38]{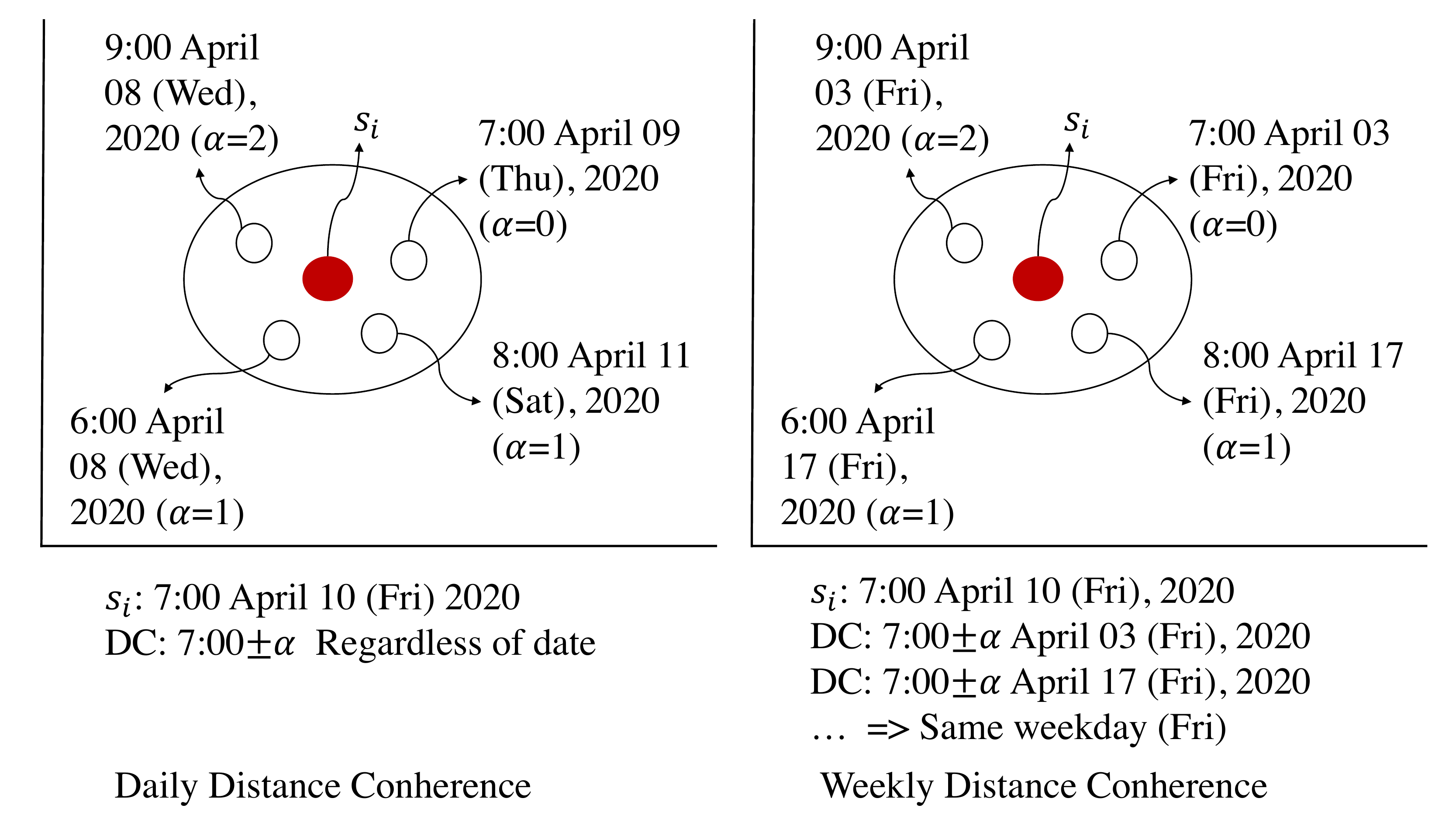}
\caption{Daily and Weekly Distance Coherence}
\label{fig:separate}
\end{figure}

This section, we describe an idea for future work based on the separation of daily and weekly distance coherences. In our current approach, for each sample $s_{i}$, the distance coherence features are calculated by grouping the other samples which have the corresponding clock hours close to the clock hour of $s_{i}$ regardless of the dates. We thus call it daily distance coherence. An example is given in the first chart of Figure~\ref{fig:separate}. The features chosen for the sample $s_{i}$ which has the timestamp at 7:00 April 10, 2020 (Friday) are calculated using the samples at 7:00$\pm \alpha$ on any date as long as they belong to the same user. 

However, another promising method may improve the accuracy or F1 score. For each sample $s_{i}$, the distance coherence features are calculated by grouping the other samples which have the clock hours close to the clock hour of $s_{i}$ on only the days that have the same day of the week. We thus call it weekly distance coherence. Look at the example in the second chart of Figure~\ref{fig:separate}, suppose $s_{i}$ occurred at 7:00 April 10, 2020 (Friday), the featured chosen for $s_{i}$ are calculated from the samples at 7:00$\pm \alpha$ on every day of Friday such as April 03, 2020 or April 17, 2020, etc. These features may reflect the lifestyle of the users that we are aiming for in this paper. For example, a worker goes to work every weekday but goes to the usual supermarket every Saturday around 10:00, a student has a training course at a usual stadium every Thursday around 15:00. These habits can be measured by the weekly distance coherence. Remark that, the weekly distance coherence features are not covered in the daily ones. Each feature is computed from the average of all the samples chosen for the main sample. Even though the set of the samples chosen for the weekly case is a subset of the set of that in the daily case, their averages are different.

\section{Conclusion}
\label{section:conclusion}

In this paper, we have shown that using the distance coherence score as the additional features can improve user authentication. We collected 107,637 GPS records including longitude, latitude, and timestamp from 348 users in Japan. The three average ensemble algorithms including RandomForest, ExtraTrees, and Bagging are applied to the classification and are evaluated using stratified $k$-fold. The experimental result showed that our approach outperforms the approach without the distance coherence in all the cases. The accuracy can reach up to 99.42\%, 99.12\%, and 99.25\% using RandomForest, ExtraTrees, and Bagging, respectively. Especially, the F1 score can be improved even 2.34\% and the false negative rate can be reduced 0.04\% using ExtraTrees. 

\appendix
\section*{Appendix}
\section{Numeric Example (for Distance Coherence Extraction)}
\label{section:appendixexample}
In this section, we give a numeric example for the distance coherence extraction in Section~\ref{section:featuredc}. Suppose the data consists of 7 samples $\{s_{1}, s_{2}, \cdots, s_{7}\}$ from 2 users $\{user1, user2\}$ as showed in Table~\ref{table:example}. We explain how to calculate the distance coherence for each sample $\{dc_{11}$, $dc_{12}$, $dc_{13}$, $dc_{21}$, $dc_{22}$, $dc_{23}$, $dc_{24}\}$. Suppose $\alpha$ (the number of distance coherence feature) is set to $\alpha = 1$. 

\begin{table}[!ht]
\centering
\caption{Numeric Example for Calculating Distance Coherence}
\begin{tabular}{c | c | c | c | c | c}
\textbf{SampleID} & \textbf{User/Class} & \textbf{Timestamp} & \textbf{Longitude} & \textbf{Latitude} & \textbf{Distance Coherence} \\
\hline
1 & user1 & 2020/01/16 10:55 & $lon_{11}$ & $lat_{11}$ & $dc_{11}$ \\
2 & user1 & 2020/01/17 11:55 & $lon_{12}$ & $lat_{12}$ & $dc_{12}$ \\
3 & user1 & 2020/01/17 12:50 & $lon_{13}$ & $lat_{13}$ & $dc_{13}$ \\
\hline
4 & user2 & 2020/01/16 21:30 & $lon_{21}$ & $lat_{21}$ & $dc_{21}$ \\
5 & user2 & 2020/01/17 22:10 & $lon_{22}$ & $lat_{22}$ & $dc_{22}$ \\
6 & user2 & 2020/01/18 21:45 & $lon_{23}$ & $lat_{23}$ & $dc_{23}$ \\
7 & user2 & 2020/01/19 20:10 & $lon_{24}$ & $lat_{24}$ & $dc_{25}$ \\
\end{tabular}
\label{table:example}
\end{table}

\begin{itemize}
\item For $s_{1}$, the hour extracted from the timestamp is $hour(s_{1}) = 10$. We find all the samples $s_{i}$ that belong to the same class ($user1$) and have $hour(s_{i})$ such that $(hour(s_{1}) - \alpha) \leq hour(s_{i}) \leq (hour(s_{1}) + \alpha)$ regardless of the date and the second. Only $s_{2}$ satisfies the conditions (i.e., $hour(s_{2}) = 11$). Thus: 
\begin{equation}
dc_{11} = \sqrt[2]{(lon_{11} - lon_{12})^{2} + (lat_{11}-lat_{12})^{2}} 
\end{equation}

\item For $s_{2}$, $hour(s_{2}) = 11$. The samples $s_{i}$ from $user1$ that satisfy $(hour(s_{2}) - \alpha) \leq hour(s_{i}) \leq (hour(s_{2}) + \alpha)$ consist of $s_{1}$ and $s_{3}$ ($hour(s_{1}) = 10, hour(s_{3}) = 12$). Thus:
\begin{equation}
dc_{12} = \sqrt[2]{(lon_{12} - \frac{lon_{11}+lon_{13}}{2})^{2} + (lat_{12}-\frac{lat_{11}+lat_{13}}{2})^{2}} 
\end{equation}

\item For $s_{3}$, $hour(s_{3}) = 12$. The sample $s_{i}$ from $user1$ that satisfies $(hour(s_{3}) - \alpha) \leq hour(s_{i}) \leq hour(s_{3}) + \alpha)$ is only $s_{2}$ ($hour(s_{2}) = 11$). Thus:
\begin{equation}
dc_{13} = \sqrt[2]{(lon_{13} - lon_{12})^{2} + (lat_{13}-lat_{12})^{2}} 
\end{equation}

\item For $s_{4}$, $hour(s_{4}) = 21$.  The samples $s_{i}$ from $user2$ that satisfy $(hour(s_{4}) - \alpha) \leq hour(s_{i}) \leq (hour(s_{4}) + \alpha)$ is $s_{5}$, $s_{6}$, and $s_{7}$ ($hour(s_{5}) = 22, hour(s_{6}) = 21, hour(s_{7}) = 20$). Thus:
\begin{equation}
dc_{21} = \sqrt[2]{(lon_{21} - \frac{lon_{22}+lon_{23}+lon_{24}}{3})^{2} + (lat_{21} - \frac{lat_{22}+lat_{23}+lat_{24}}{3})^{2}} 
\end{equation}

\item For $s_{5}$, $hour(s_{5})=22$. The samples $s_{i}$ from $user2$ that satisfy $(hour(s_{5}) - \alpha) \leq hour(s_{i}) \leq (hour(s_{5}) + \alpha)$ is $s_{4}$ and $s_{6}$ ($hour(s_{4}) = hour(s_{6}) = 21$). Thus:
\begin{equation}
dc_{22} = \sqrt[2]{(lon_{22} - \frac{lon_{21}+lon_{23}}{2})^{2} + (lat_{22} - \frac{lat_{21}+lat_{23}}{2})^{2}} 
\end{equation}

\item For $s_{6}$, $hour(s_{6})=21$. The samples $s_{i}$ from $user2$ that satisfy $(hour(s_{6}) - \alpha) \leq hour(s_{i}) \leq (hour(s_{6}) + \alpha)$ is $s_{4}$, $s_{5}$, and $s_{7}$ ($hour(s_{4}) = 21, hour(s_{5}) = 22, hour(s_{7}) = 20$). Thus:
\begin{equation}
dc_{23} = \sqrt[2]{(lon_{23} - \frac{lon_{21}+lon_{22}+lon_{24}}{3})^{2} + (lat_{23} - \frac{lat_{21}+lat_{22}+lat_{24}}{3})^{2}} 
\end{equation}

\item For $s_{7}$, $hour(s_{7})=20$. The samples $s_{i}$ from $user2$ that satisfy $(hour(s_{7}) - \alpha) \leq hour(s_{i}) \leq (hour(s_{7}) + \alpha)$ is $s_{4}$ and $s_{6}$ ($hour(s_{4}) = hour(s_{6}) = 21$). Thus:
\begin{equation}
dc_{24} = \sqrt[2]{(lon_{24} - \frac{lon_{21}+lon_{23}}{2})^{2} + (lat_{24} - \frac{lat_{21}+lat_{23}}{2})^{2}} 
\end{equation}

\end{itemize}


\begin{thebibliography}{30}










\bibitem{RandomForest2001}
Breiman L (2001) Random Forests. In: Machine Learning, 45(1):5-32.

\bibitem{ExtraTrees2006}
Geurts P, Damien E, Wehenkel L (2006) Extremely randomized trees. In: Machine Learning, 63(1):3-42.

\bibitem{Bagging2012}
Louppe G and  Geurts P (2012) Ensembles on Random Patches. In: European Conference on Machine Learning and Principles and Practice of Knowledge Discovery in Databases (ECML PKDD'12), pp. 346-361.

\bibitem{society50}
Cabinet Office, the Government of Japan, Society 5.0. Available: \url{https://www8.cao.go.jp/cstp/english/society5_0/index.html}. Latest Accessed: April 26, 2020.

\bibitem{gps2016}
Fridman L, Steven W, Rachel G, Moshe K (2016) Active Authentication on Mobile Devices via Stylometry, Application Usage, Web Browsing, and GPS Location. In: IEEE Systems Journal, 11(2):513-521.


\bibitem{gps2011}
Shi W, Yang J, Jiang Y, Yang F, Xiong Y (2011) SenGuard: passive user identification on smartphones using multiple sensors. In: IEEE 7th International Conference on Wireless and Mobile Computing, Networking and Communications (WiMob'11), pp. 141-148. 

\bibitem{thaodbsec}
Thao T.P., Irvan M., Kobayashi R., Yamaguchi R.S., Nakata T. (2020) Self-enhancing GPS-Based Authentication Using Corresponding Address. In: Data and Applications Security and Privacy XXXIV (DBSec'20), Lecture Notes in Computer Science, vol. 12122. Springer, Cham, pp. pp 333--344. DOI: \url{https://doi.org/10.1007/978-3-030-49669-2_19}

\bibitem{gpsMULEA2019}
Alejandro A, Aythami M, Vera-Rodriguez R, Julian F, Ruben T (2019) MultiLock: Mobile Active Authentication based on Multiple Biometric and Behavioral Patterns. In: International Workshop on Multimodal Understanding and Learning for Embodied Applications (MULEA'19), pp. 53-59.

\bibitem{gpsuspatent2018}
Aaron B, Christopher D, Barry G, David K (2018) System and method for real world biometric analytics through the use of a multimodal biometric analytic wallet. In: US patent, US20180276362A1. Available: \url{https://patents.google.com/patent/US20100050253}. Latest Accessed: April 26, 2020.

\bibitem{gpsacm2018}
Valentin R, Catherine T,  Sourav B, Nicholas L, Cecilia M, Mahesh M, Fahim K (2018) Multimodal Deep Learning for Activity and Context Recognition. In: Publication:Proceedings of the ACM on Interactive, Mobile, Wearable and Ubiquitous Technologies (IMWUT'18), article no. 157, DOI: \url{https://doi.org/10.1145/3161174}. Latest Accessed: April 26, 2020.

\bibitem{introsoup2013}
Dirk B,  Shu L, Mitch K, Aaron S, Charles C, John D (2013) Modifying smartphone user locking behavior. In: 9th Symposium on Usable Privacy and Security (SOUPS'13), article no. 10, pp. 1-14, DOI: \url{https://doi.org/10.1145/2501604.2501614}. Latest Accessed: April 26, 2020.

\bibitem{introccs2014}
Egelman S, Jain S, Portnoff R, Liao K, Consolvo S, Wagner D (2014) Are you ready to lock?. In: 21st ACM SIGSAC Conference on Computer and Communications Security (CCS'14), pp. 750-761, DOI: \url{https://doi.org/10.1145/2660267.2660273}. Latest Accessed: April 26, 2020.

\bibitem{introsoup2014}
Harbach M, Zezschwitz E, Fichtner A, Luca A, Smith M (2014) It's a hard lock life: A field study of smartphone (un) locking behavior and risk perception. In: 10th USENIX Conference on Usable Privacy and Security (SOUP'14), pp. 213-230.

\bibitem{gpsupal2016}
Upal M, Sayantan S, Vishal P, Rama C (2016) Active user authentication for smartphones: A challenge data set and benchmark results. In: 8th International Conference on Biometrics Theory, Applications and Systems (BTAS'16), DOI: 10.1109/BTAS.2016.7791155.

\bibitem{hashcat}
Hashcat - Advanced Password Recovery. Available: \url{https://hashcat.net/hashcat/}. Latest Accessed: April 27, 2020.

\bibitem{johnripper}
John the Ripper password cracker. Available: \url{https://www.openwall.com/john/}.  Latest Accessed: April 27, 2020.

\bibitem{iris}
Marsicoa MD, Michele N, Daniel R, Wechsler H (2015) Mobile Iris Challenge Evaluation (MICHE)-I, biometric iris dataset and protocols. In: Pattern Recognition Letters, vol. 57, pp. 17-23, Elsevier, DOI: \url{https://doi.org/10.1016/j.patrec.2015.02.009}. Latest Accessed: April 27, 2020.

\bibitem{irisoriginal}
Venugopalan S and Savvides M (2011) How to Generate Spoofed Irises From an Iris Code Template. In: IEEE Transactions on Information Forensics and Security, 6(2):385-395, DOI: \url{https://doi.org/10.1109/TIFS.2011.2108288}. Latest Accessed: April 27, 2020.

\bibitem{contactlens}
Kevin W. Bowyer and James S. Doyle (2014) Cosmetic Contact Lenses and Iris Recognition Spoofing. In: Computer, 47(5):96-98, DOI: 10.1109/MC.2014.118.

\bibitem{contactlens1}
Daksha Y, Naman K, James SD, Richa S, Mayank V, Kevin WB (2014) Unraveling the Effect of Textured Contact Lenses on Iris Recognition. In: IEEE Transactions on Information Forensics and Security, 9(5):851-862, DOI: 10.1109/TIFS.2014.2313025.

\bibitem{facialvideo}
Chingovska I, Anjos A, Marcel S (2012) On the effectiveness of local binary patterns in face anti-spoofing. In: IEEE International Conference of Biometrics Special Interest Group (BIOSIG'12). 

\bibitem{facial3D}
Erdogmus N and Marcel S (2013) Spoofing in 2D face recognition with 3D masks and anti-spoofing with Kinect. In: IEEE 6th International Conference on Biometrics: Theory Applications and Systems (VISAPP'13), pp. 1-6.

\bibitem{fingerprintingsp}
Anthony R (2009) System for and method of securing fingerprint biometric systems against fake-finger spoofing. US Patent US7505613B2. Available: \url{https://patents.google.com/patent/US7505613B2/en}. Latest Accessed: April 27, 2020.

\bibitem{combine}
David M, Giovani C, Allan P, William RS, Helio P, Alexandre XF, Anderson R (2015) Deep Representations for Iris, Face, and Fingerprint Spoofing Detection. In: IEEE Transactions on Information Forensics and Security, 10(4):864-879.

\bibitem{smudge}
Adam JA, Katherine G, Evan M, Matt Blaze, Jonathan MS (2010) Smudge attacks on smartphone touch screens. In: 4th USENIX conference on Offensive technologies (WOOT'10), pp. 1-7.

\bibitem{chi17}
Malin E, Mohamed K, Emanuel VZ, Heinrich H, Florian A (2017) Understanding Shoulder Surfing in the Wild: Stories from Users and Observers. In: ACM CHI Conference on Human Factors in Computing Systems, pp. 4254-4265, DOI: \url{https://doi.org/10.1145/3025453.3025636}.

\bibitem{other1}
Kaczmarek T, Ercan O, Gene T (2018) Assentication: User Deauthentication and Lunchtime Attack Mitigation with Seated Posture Biometric. In: 16th International Conference on Applied Cryptography and Network Security (ACNS'18), pp. 616-633.

\bibitem{other2}
Ivan M, Kasper R, Marc R, Gene T (2017) Authentication U­sing Pulse-Response Biometrics. In: Communications of the ACM, 60(2):108-115. (A full version of this paper was presented at the Network and Distributed System Security (NDSS) Symposium 2014).

\bibitem{adversarialclassifier}
Nilesh D, Pedro D, Mausam, Sumit KS, Deepak V (2004) Adversarial classification. In: 10th ACM SIGKDD international conference on Knowledge discovery and data mining (KDD'04), pp. 99-108, DOI: \url{https://doi.org/10.1145/1014052.1014066}. 

\bibitem{othereyemovement}
Simon E, Kasper BR, Vincent L, Ivan M (2015) Preventing Lunchtime Attacks: Fighting Insider Threats With Eye Movement Biometrics. In: 22nd Annual Network and Distributed System Security Symposium (NDSS'15). 
 
\bibitem{signal1}
Louis W, Komeili M, Hatzinakos D (2016) Continuous authentication using one-dimensional multi-resolution local binary patterns (1dmrlbp) in ecg biometrics. In: IEEE Transactions on Information Forensics and Security, 11(12):2818-2832.

\bibitem{ACHI}
Thao TP, Takahashi M, Shigeta N, Irvan M, Nakata T, and Yamaguchi RS, ``Human Factors in Exhaustion and Stress of Japanese Nursery Teachers: Evidence from Regression Model on A Novel Dataset''. In: 13th International Conference on Advances in Computer-Human Interactions (ACHI'20), pp. 124--129. 

\bibitem{signal2}
Alejandro R, Stephen D, Ivan C, Giulio R (2008) STARFAST: a Wireless Wearable EEG/ECG Biometric System based on the ENOBIO Sensor. In: International Workshop on Wearable Mycro and Nanosystems for Personalised Health (pHealth'08). 

\bibitem{Google1}
N. Neverova, W. Christian, G. Lacey, L. Fridman, D. Chandra, B. Barbello, and G. Taylor. ``Learning Human Identity From Motion Patterns''. In: \emph{IEEE Access }, vol. 4. pp. 1810-1820, 2016.
 
\end{thebibliography}
\end{document}